\documentclass[aip,amsmath,amssymb,reprint,]{revtex4-2}

\usepackage{graphicx}
\usepackage{dcolumn}
\usepackage{bm}
\usepackage{epstopdf}
\usepackage{xcolor}
\usepackage{textcomp}
\usepackage{soul}
\usepackage{csquotes}
\usepackage{amsbsy}
\usepackage{mathrsfs} 
\usepackage{amsmath}
\usepackage{bigints} 
\usepackage{amsthm,amssymb} 
\usepackage[utf8]{inputenc}
\usepackage[english]{babel}

\usepackage[utf8]{inputenc}
\usepackage[T1]{fontenc}
\usepackage{mathptmx}
\usepackage{xcolor}
\usepackage[shortlabels]{enumitem}
\usepackage{float}
\usepackage{mathrsfs,bigints,mathtools,dsfont}
\usepackage[colorlinks=true,linkcolor=blue,citecolor=blue]{hyperref}%

\begin{document}

	\title{Intralayer and interlayer synchronization in multiplex network with higher-order interactions}
	
	\author{Md Sayeed Anwar}
		\affiliation{%
		Physics and Applied Mathematics Unit, Indian Statistical Institute, 203 B. T. Road, Kolkata 700108, India}
	
	\author{Dibakar Ghosh}
	\email{diba.ghosh@gmail.com}
	\affiliation{Physics and Applied Mathematics Unit, Indian Statistical Institute, 203 B. T. Road, Kolkata 700108, India}	
	

\begin{abstract}
 Recent developments in complex systems have witnessed that many real-world scenarios, successfully represented
 as networks are not  always restricted to binary interactions but often include higher-order interactions among the nodes. These beyond pairwise interactions are preferably modeled by hypergraphs, where hyperedges represent higher-order interactions between a set of nodes. In this work, we consider a multiplex network where the intralayer connections are represented by hypergraphs, called multiplex hypergraph. The hypergraph is constructed by mapping the maximal cliques of a scale-free network to hyperedges of suitable sizes. We investigate the intralayer and interlayer synchronization of such multiplex structures. Our study unveils that the intralayer synchronization appreciably enhances when the higher-order structure is taken into consideration inspite of only pairwise connections. We derive the necessary condition for stable synchronization states by the master stability function approach, which perfectly agrees with the numerical results. We also explore the robustness of interlayer synchronization and find that for the multiplex structures with many-body interaction, the interlayer synchronization is more persistent than multiplex networks with pairwise interaction.   	
\end{abstract}

\maketitle

\begin{quotation}
   In recent years, the accessibility to further detailed and massive datasets reveals that from social to biological systems, real-world interactions generally take place between a group of more than two agents at an instance. This situation is not appropriately explained by conventional two-body interactions and emphasizes the necessity for higher-order interactions. Including the higher-order interactions in the multiplex network can considerably influence collective phenomenons that emerge through interactions between dynamical units. By investigating the synchronization phenomena, particularly intralayer and interlayer synchronization on multiplex networks with higher-order dynamics, we perceive that intralayer synchronization greatly enhances and interlayer synchronization is more robust compared to the pairwise scenario. We derive the necessary conditions for the existence and stability of the synchronization states on multiplex networks with higher-order interactions and examine its vitality by means of numerical simulations and spectral interpretation.      
\end{quotation}

\section{Introduction}
Over the past 20 years, the theory of complex networks has been able to successfully narrate the collective behaviors arising from  a large number of interacting dynamical units in many real-world systems.\cite{network1,albert_barabasi,boccaletti2006complex} However, the pairwise interactions schematized by edges and modeled by networks are not always capable of describing underlying connections between generic units of many circumstances. From functional \cite{functional_brain1,functional_brain2} and structural brain networks \cite{structural_brain} to protein interaction networks,\cite{protein} to semantic networks, \cite{semantic} random walks \cite{random_walk,random_pre}, collaboration graphs \cite{coauthor1,coauthor2}, epidemiology \cite{contagion1,contagion2} and ecological communities\cite{ecological1,ecological2} there are a lot of situations which simply can not be described only through the perspective of pairwise interactions \cite{majhi2022}. In all these examples, interactions are not limited to pairwise but occur in groups of arbitrary sizes.\cite{beyond_pairwise,skardal2021higher,bick2021higher} So, one needs a generalization of graphs to study these systems properly. Hypergraphs \cite{hypergraph1,hypergraph2,hypergraph3,hypergraph4} and simplicial complexes \cite{simplicial1,simplicial2,simplicial3}are the natural generalizations of graphs to describe the interactions beyond pairwise. Hypergraphs materialized as higher-order interactions, represented by a collection of any number of nodes assembled in sets, called hyperedges among generic agents. On the other hand, simplicial complexes represent higher-order interactions following an additional constraint about including all lower-order interactions, i.e., for instance, a three-body interaction requires the existence of all pairwise connections associated with the same triangle. So, the hypergraphs is a more general framework to address many-body interactions between generic nodes.  

\par  Synchronization,\cite{synchronization1,synchronization2,synchronization3} one of the collective phenomena ubiquitous in natural systems, emerges through the adjustment of evolving dynamical units. Recently, the study of synchronization in networks with higher-order interaction has received much attention among researchers from various fields. In this regard, most of the studies on synchronization involving higher-order structures consider simplicial complexes \cite{simplicialsync2,simplicialsync3,simplicialsync4,simplicialsync6} to represent non-pairwise interactions due to their simple geometrical representation. Very few have taken hypergraphs\cite{de2021phase,banerjee2020synchronization,bohle2021coupled,schaub1,schaub2} into consideration. The extension of complete synchronization to $p$- uniform hypergraphs have been studied in Ref.~\onlinecite{krawiecki2014chaotic}. A new perception for the master stability function of steady states and synchronization in chemical hypergraphs has been reported by Mulas et al. in Ref.~\onlinecite{chem_hypergraph_msf}. Wu et al. \cite{wu2014synchronization} have derived analytical criteria for the synchronization of Chua oscillators in q-hypergraphs. A different approach to extend the master stability function formalism based on a new hypergraph Laplacian is introduced in Ref.~\onlinecite{carletti2020dynamical}. 

\par On the other hand, multilayer
networks,\cite{boccaletti2014structure,kivela2014multilayer} for instance, multiplex networks, develop a very relevant network configuration to characterize real-world scenarios where the activity of a network affects other networks. This type of network consists of more than one layer, where interactions in each layer may differ from other layers. As an analogy, mobility networks,\cite{mobility} power-grid networks,\cite{power} air transportation networks,\cite{air} social networks,\cite{social} neuronal networks,\cite{neuronal} ecological networks\cite{ecology} are well illustrated by such frameworks. Various collective synchronization states emerge due to interactions in multilayered systems, namely intralayer\cite{intra1,intra2} and interlayer\cite{inter1,inter2} synchronization, cluster synchronization,\cite{cluster1,cluster2} explosive synchronization,\cite{explosive1,explosive2} antiphase synchronization,\cite{antiphase} relay synchronization \cite{relay1,relay2}and chimera states.\cite{chimera1,chimera2} 

\par In the previous studies on multiplex networks, the connections within layers are limited to only two-body interactions, i.e., represented by graphs through pairwise links. However, the pairwise representation of layers in multiplex networks is not always able to capture the interaction among agents accurately. For instance, in the multiplex representation of social networks with layers representing the interconnections between groups of people through different social media applications, the group chat between a group of friends where everyone can connect to each other without having individual pairwise communication is not possible to delineate with only the pairwise connections. To describe these types of situations higher-order network structures, i.e., hypergraphs are far more appropriate. Motivated by this, here we desert the curb to stick with the pairwise representation of multiplex networks and consider many-body interactions through hypergraphs in layers of the multiplex network. A multiplex network is generally constructed by two different types of connections: intralayer and interlayer links. Intralayer links correspond to the connection of units within the layers, and the latter indicates interconnection between each node of a layer and all its replicas in other layers. In this work, we consider the first type of interaction as higher-order, constructed by hypergraphs, and the second represents pairwise interaction between counterpart nodes. This type of multiplex stricture is termed as multiplex hypergraph. As the previous studies on the synchronization in multiplex network structure, only the pairwise interaction between the units in the layers is considered, here we concentrate on the study of synchronization phenomena in the multiplex hypergraph, particularly the intralayer and interlayer synchronization. We derive the necessary condition for stability of the two aforesaid synchronization states in the multiplex hypergraph analytically using the master stability function \cite{msf} (MSF) scheme. The analysis shows that the synchronization of layers with multiplex hypergraphs enhances compared to multiplex networks constructed by only pairwise interactions. For the acute understanding of our derived result, we also study the spectral properties of the layer networks in both scenarios: binary and higher-order interactions. Further, we broaden our study by investigating the robustness of interlayer synchrony against a progressive detachment of links between replica nodes.

\par The present article is structured as follows: Section II is assigned for a brief review of hypergraphs and multiplex hypergraphs. In Sec. III, we introduce the model corresponding to multiplex hypergraphs. The variation of synchronization error with respect to coupling strengths is explored thoroughly in Sec. IV. Sec. IV A, represents the linear stability analysis of synchronization states, and Sec. IV B, represents the spectral analysis to support our analytical results. Next, we study the persistence of interlayer synchronization in Sec. IV C. Finally, we sum up our results and conclude in Sce. V.      

 \begin{figure*}[ht]
	\centerline{\includegraphics[scale=0.35]{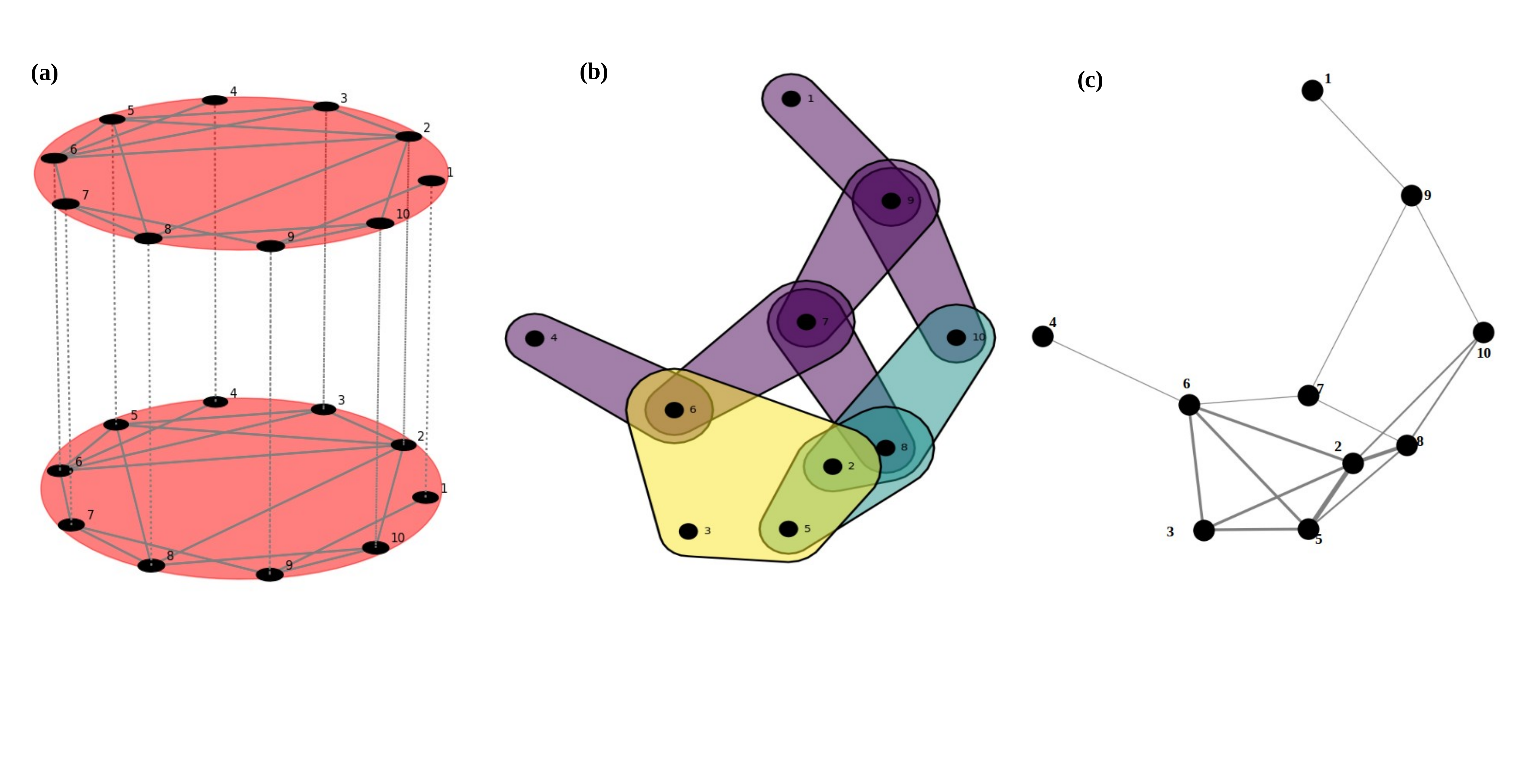}}
		\caption{{ \bf Schematic diagram of multiplex hypergraphs and corresponding networks.} The pairwise multiplex network is reported in (a). The connection mechanism of layers corresponds to multiplex hypergraph is depicted in (b). The hyperedges are colored according to their sizes (yellow for size 4, green for size 3, and violet for size 2). The hypergraph is constructed by mapping the maximal cliques of the pairwise multiplex network to hyperedges of suitable sizes. The equivalent weighted network of  the hypergraph is presented in (c). The width of edges corresponds to their respective weight are obtained from Eq. \eqref{eq.2}.} 
	\label{fig1}
\end{figure*}

\section{Hypergraphs and Multiplex Hypergraphs} A hypergraph is a dyad $\mathscr{H}=(\mathscr{V}, \mathscr{E})$, where $\mathscr{V}=\{1, 2, \cdots, N\}$ is the collection of $N$ nodes and $\mathscr{E}=\{E_1, E_2, \cdots, E_M\}$ characterizes collection of $M$ hyperedges. The hyperedges $E_h$, for all $h \in \{ 1, 2, \cdots, M \}$ are a collection of nodes that represents interactions of arbitrary order among the units, i.e., $E_h \subset \mathscr{V}$. When $E_h=(i, j )$, that is, $| E_h|= 2$ then the hyperedge is just a usual edge representing pairwise interaction among $i$ and $j$. If the size of all hyperedges is 2 then the hypergraph becomes a pairwise interaction graph.
\par To describe how the nodes are allocated among the hyperedges, we can define the incidence matrix of the hypergraph $\mathscr{I}=\{\mathscr{I}_{ih}\}$ as 
  \begin{equation}{\label{eq.1}}
  \begin{array}{l}
   \mathscr{I}_{ih}= \begin{cases}
            1, &  {i}\in E_{h} \\
            0, &  \mbox{otherwise}.
   \end{cases}
  \end{array}
  \end{equation}         

 The above matrix defines $N \times N$ adjacency matrix $\mathscr{A}={\mathscr{I}\mathscr{I}^T}$ by setting the diagonal elements to be zero, whose entries represent the number of hyperedges shared by two adjacent nodes and the $M \times M$ hyperedge matrix $\mathscr{C}=\mathscr{I}^T\mathscr{I}$, whose entries $\mathscr{C}_{h_1h_2}$ gives the number of nodes in $E_{h_1}\cap E_{h_2}$. The corresponding Laplacian matrix \cite{chem_hypergraph_msf,laplacian1} $\mathscr{L}$, whose $(i,j)$ term is given by $\mathscr{L}_{ij}=d_{i}\delta_{ij}-\mathscr{A}_{ij}$, where $d_{i}=\sum\limits_{j=1}^{N}\mathscr{A}_{ij}$ counts the number of hyperedges passes through the node $i$ and $\delta_{ij}$ represents kronecker delta. This is a natural generalization of the pairwise Laplacian matrix. Nevertheless, the Laplacian matrix lacks in describing the entire higher-order structure encrypted in the hypergraph. Specifically, the sizes of the hyperedges are overlooked. To go beyond this bound, a new Laplacian matrix is introduced in Ref.~\onlinecite{random_pre}, to study the random walk process on hypergraphs based on the fact that exchanges among nodes belonging to the same hyperedge are more favored than nodes linked to other hyperedges. This newly defined Laplacian matrix takes into account both the sizes and numbers of hyperedges incident to a node by assigning weight to the edges according to the size of hyperedges. In particular, the hypergraph is transformed into an equivalent weighted network whose weights have been defined to represent the higher-order structure encrypted in the hypergraph. The weighted adjacency matrix $\mathscr{W}$ is given by,
 \begin{equation}{\label{eq.2}}
 \begin{array}{l}
 \mathscr{W}_{ij}=\begin{cases}

 \sum\limits_{h} (\mathscr{C}_{hh}-1)\mathscr{I}_{ih}\mathscr{I}_{jh}=(\mathscr{I}\hat{\mathscr{C}}\mathscr{I}^{T})_{ij}-\mathscr{A}_{ij}, & i \ne j \\
 0,  ~~~~~~ \mbox{otherwise},
 \end{cases}
 \end{array}
 \end{equation}
 where $\hat{\mathscr{C}}$ is a diagonal matrix whose nonzero entries are same as the diagonal elements of $\mathscr{C}$. It is clear from its definition that the weighted network considers both the sizes and numbers of hyperedges incident to a node. The corresponding Laplacian matrix \cite{random_pre,carletti2020dynamical} for equivalent weighted network of the hypergraph is given by 
 \begin{equation}{\label{eq.3}}
 \begin{array}{l}
 L^{H}=D-\mathscr{W},
 \end{array}
 \end{equation}
 where $D$ is the matrix with diagonal entries $\sum\limits_{j=1}^{N}\mathscr{W}_{ij}$ and zero otherwise. When the size of each hyperedge is 2, then the Laplacian matrix converts into standard graph Laplacian $L$. On this note, it is necessary to emphasize that the dynamics represented on the weighted network are analogous with the dynamics on the hypergraph.\cite{random_walk} Hence we can use the existing tools of the network to analyze various phenomena between interacting units described through hypergraphs. Notably, dynamical systems on hypergraphs can be studied with $N\times N$ matrices to avoid difficulties, unlike simplicial complexes, where higher-order tensors are involved. 
 
 \par Now, we can define a multiplex hypergraph $\mathscr{M}=(\mathscr{H},\mathscr{P})$, where $\mathscr{H}=\{\mathscr{H}_{\alpha}=(\mathscr{V}_{\alpha},\mathscr{E}_{\alpha}):\alpha \in \{1, 2, \cdots, L\}\}$ is an ensemble of hypergraphs, each portraying a layer with a fixed number of nodes in each layer, i.e., $|\mathscr{V}_{\alpha}|=N$ for all $\alpha$ and $\mathscr{P}$ is the collection of links between each node of a layer and all its replicas in the other layers. The elements of $\mathscr{E}_{\alpha}$ are intralayer connections representing higher-order structures via hyperedges of several sizes, and the elements of $\mathscr{P}$ are pairwise interlayer connections. The multiplex hypergraph is a generalization of the pairwise multiplex network that only considers pairwise connections between nodes in each layer. 

\par From a given hypergraph, it is possible to create a suitable pairwise network by mapping the units associated to a hyperedge into a clique of appropriate size. This mapping is invertible if the hypergraph contains only simple hyperedges, i.e., given a pairwise network, one can construct a hypergraph by converting a maximal $p$ clique to a hyperedge of size $p$\cite{random_pre}.

 \par The schematic diagram (Figure \ref{fig1}(a)) depicts a pairwise multiplex network.  The corresponding multiplex hypergraph is constructed by mapping the maximal $p$ cliques of each layer into hyperedges of size $p$. Figure \ref{fig1}(b) represents a layer of the multiplex hypergraph, and its corresponding equivalent weighted network is displayed in Figure \ref{fig1}(c).

\par 
 Once again, we emphasize that the equivalent weighted network represents the same dynamics as the hypergraph. For this reason, we present our multiplex hypergraph  through weighted one to easily handle the dynamics on multiplex hypergraph with the existing tool of network theory.  
  
\section{Mathematical structure of multiplex hypergraph} 

We consider a two-layer multiplex hypergraph, each layer contains $N$ nodes of $d$-dimensional identical oscillators interacting diffusively through hyperedges of various sizes. The layer-wise dynamics of the multiplex hypergraph are expressed in terms of the state vectors ${\bf X_{l}} = ({\bf x}_{l,1}, {\bf x}_{l,2}, \dots, {\bf x}_{l,N})$, where ${\bf x}_{l,i} \in \mathbb{R}^d$ and $i = 1, 2, \cdots, N; ~ l=1,2$. We can write the equation of motion of each dynamical unit by the following system of equations, 
\begin{equation}{\label{eq.4}}
\begin{array}{lcl}
	
	\mbox{Layer-1:} \\\\
	\dot{\bf x}_{1,i} = F({\bf x}_{1,i})+\sigma\sum\limits_{h : i \in E_{h}, j \in E_{h} }\sum\limits_{j \ne i}({\mathscr{C}_{hh}}-1) [G({\bf x}_{1,j})-G({\bf x}_{1,i})] \\\\~~~~~~~~~~~~~+\lambda [ H({\bf x}_{2,i}) - H({\bf x}_{1,i})], \\\\
	
	\mbox{Layer-2:} \\\\
	\dot{\bf x}_{2,i} = F({\bf x}_{2,i})+\sigma\sum\limits_{h : i \in E_{h}, j \in E_{h} }\sum\limits_{j \ne i}({\mathscr{C}_{hh}}-1) [G({\bf x}_{2,j})-G({\bf x}_{2,i})] \\\\~~~~~~~~~~~~~+\lambda [ H({\bf x}_{1,i}) - H({\bf x}_{2,i})], \\
\end{array}
\end{equation}   
where $F:\mathbb{R}^d\to\mathbb{R}^d $ describes the local dynamics of uncoupled oscillators, $G:\mathbb{R}^d \times \mathbb{R}^d \to\mathbb{R}^d $, and $H:\mathbb{R}^d \times \mathbb{R}^d \to\mathbb{R}^d $ represent intralayer, and interlayer coupling functions, respectively. $\sigma$ and $\lambda$ are real valued parameters describing intralayer and interlayer coupling strengths. The size of the hyperedge $E_{h}$ is denoted by the elements $\mathscr{C}_{hh}$ of the matrix $\mathscr{C}$ and the $-1$ term accounts for the fact that $i$ and $j$ must be different. Recollecting the definition of incidence matrix $\mathscr{I}$, we can rewrite the above equation corresponding to Layer-1 as

\begin{equation}{\label{eq.5}}
	\begin{array}{lcl}
		\dot{\bf x}_{1,i} = F({\bf x}_{1,i})+\sigma\sum\limits_{h=1}^{M}\sum\limits_{j=1}^{N} \mathscr{I}_{ih} \mathscr{I}_{jh} ({\mathscr{C}_{hh}}-1) [G({\bf x}_{1,j})-G({\bf x}_{1,i})] \\~~~~~~~~~~~~~+\lambda [ H({\bf x}_{2,i}) - H({\bf x}_{1,i})] \\
		
		~~~~~~~= F({\bf x}_{1,i})+\sigma\sum\limits_{j=1}^{N}{\mathscr{W}_{ij}} [G({\bf x}_{1,j})-G({\bf x}_{1,i})] \\~~~~~~~~~~~~~+\lambda [ H({\bf x}_{2,i}) - H({\bf x}_{1,i})] \\
		
		~~~~~~~=  F({\bf x}_{1,i})-\sigma\sum\limits_{j=1}^{N}{{L}_{ij}^{H}} G({\bf x}_{1,j})+\lambda [ H({\bf x}_{2,i}) - H({\bf x}_{1,i})],
	\end{array}
\end{equation}
where we have used the definitions of hypergraph equivalent weighted adjacency matrix $(\mathscr{W})_{N\times N}$ given by Eq. (\ref{eq.2}) and Laplacian matrix given by Eq. (\ref{eq.3}). Similarly the equation of motion of Layer-2 can be simplified as 

\begin{equation}{\label{eq.6}}
\begin{array}{lcl}
\dot{\bf x}_{2,i} = F({\bf x}_{2,i})-\sigma\sum\limits_{j=1}^{N}{{L}_{ij}^{H}} G({\bf x}_{2,j}) + \lambda [ H({\bf x}_{1,i}) - H({\bf x}_{2,i})].
\end{array}
\end{equation}

We assume the individual node dynamics of each layer as identical chaotic R\"{o}ssler oscillator given by,

\begin{equation}\label{eq.7}
F({\bf x}_{l,i}) = 
\begin{pmatrix}
- y_{l,i} - z_{l,i} \\
x_{l,i} + 0.2y_{l,i} \\
0.2 + z_{l,i}(x_{l,i}-5.7)
\end{pmatrix}.
\end{equation} 
 The intralayer and interlayer coupling functions are taken as $G({\bf x})=[0, y, 0]^{tr}$ and $H({\bf x})=[0, y, 0]^{tr}$, where $[\hspace{2pt} ]^{tr}$ denotes transpose of a matrix. These coupling through y-variable generate a class-II MSF. \cite{boccaletti2006complex} 
 
 \par
\begin{figure}[ht]
	\centerline{\includegraphics[scale=0.08]{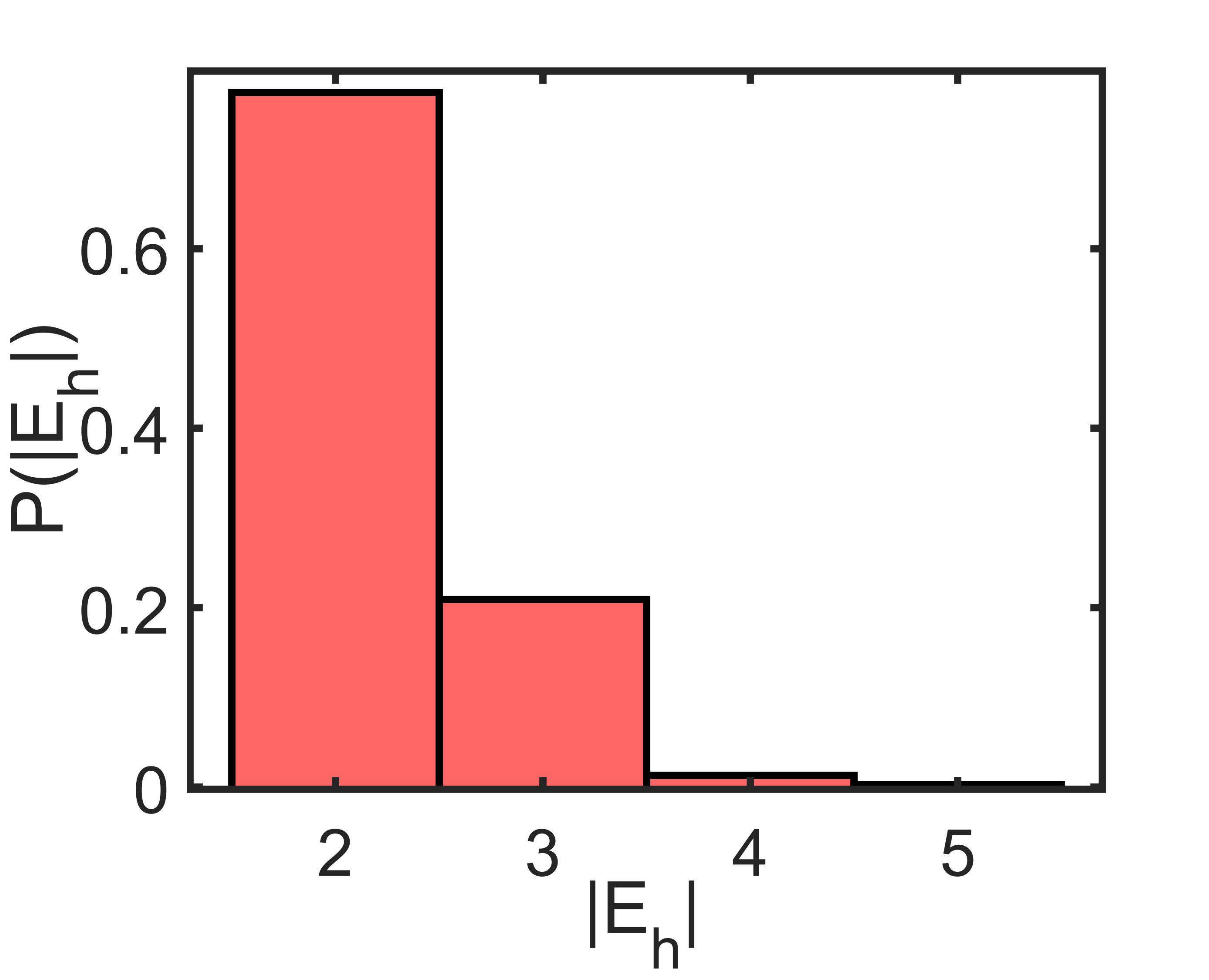}}
	
	\caption{ {\bf Hyperedges size distribution.} The distribution of sizes of hyperedges of the hypergraph is shown. The hypergraph is constructed by mapping the maximal cliques to hyperedges of suitable sizes from a scale-free network with $N=500$ nodes and $m=3$ new links added per growth step through preferential attachment.} 
	\label{fig2}
\end{figure}

  To elucidate our result, we consider the intralayer connection topology of each layer for the case of the pairwise multiplex network as a scale-free network with $N=500$ nodes and $m = 3$ links added per growth step, tied up by preferential attachment following the model proposed in Ref.~\onlinecite{albert_barabasi}. From the pairwise multiplex network, layers of the corresponding multiplex hypergraph are obtained by converting all the maximal-$p$ cliques of the pairwise network to hyperedges of sizes $p$, i.e., if there is a clique of size $p$ in the pairwise network which is not a subset of $p+1$ clique, then the corresponding $p$ clique will be converted into a hyperedge of size $p$ to construct the hypergraph. The distribution of the hyperedges of the corresponding hypergraph associated with the layers of the multiplex hypergraph is delineated in Figure \ref{fig2}.

\begin{figure*}[ht]
\centerline{\includegraphics[scale=0.4]{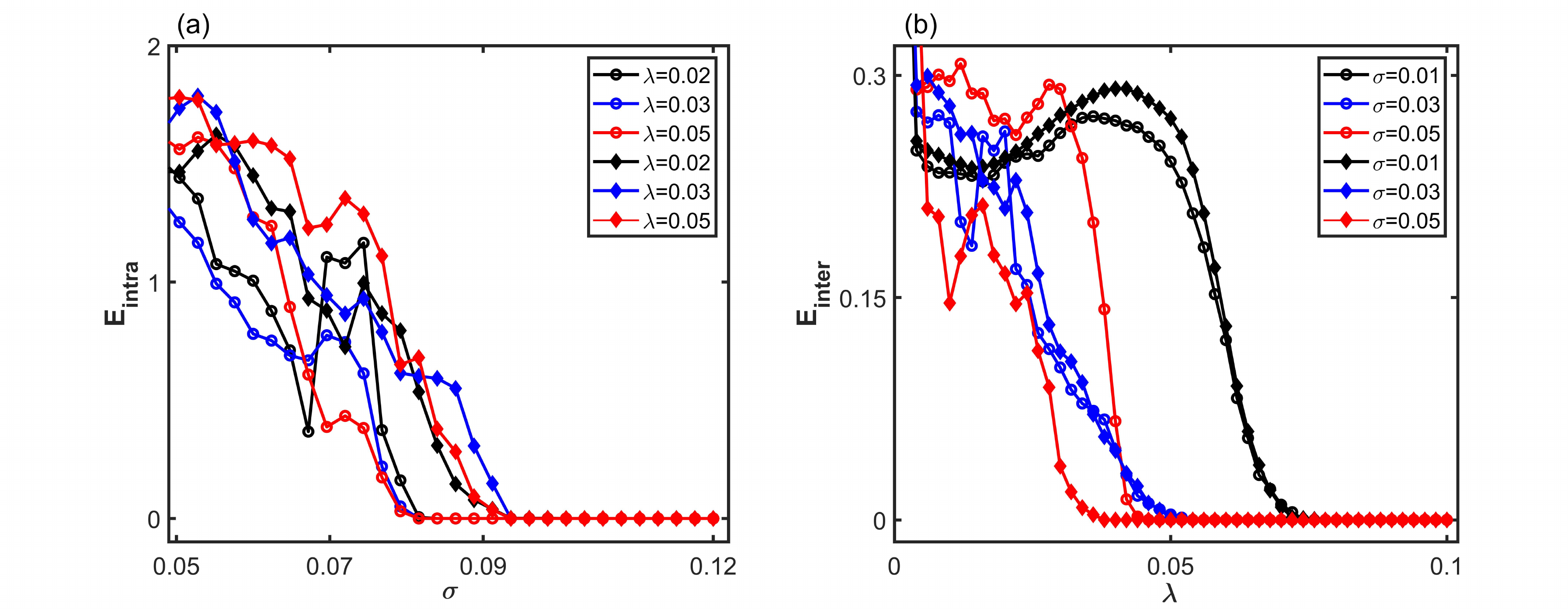}}
	\caption{{\bf Synchronization errors with respect to coupling strengths.} (a) $E_{intra}$, intralayer synchronization error as a function of $\sigma$ for various values of $\lambda$ (in legend ). (b) $E_{inter}$, interlayer synchronization error as a function of $\lambda$ for several values of $\sigma$ ( in legend ). Solid diamonds are for pairwise multiplex network while void circles corresponds to multiplex hypergraph. } 
	\label{fig3}
\end{figure*}

\par In the following sections, the intralayer and interlayer synchronization in the multiplex hypergraph (\ref{eq.4}) are scrutinized, and the results are compared with  the pairwise multiplex network, where connections between the nodes of a particular layer are represented only by pairwise links.

\section{Results}

 To investigate the intralayer and interlayer synchronization, we define the synchronization errors as,
  
\begin{equation}{\label{eq.8}}
\begin{array}{lcl}
E_{intra} = \frac{E_{intra_1} + E_{intra_2}}{2},
\end{array} 
\end{equation}
with
$$E_{intra_1} = \lim\limits_{T\to \infty} \frac{1}{T} \int_{t_{trans}}^{t_{trans}+T} \sum\limits_{j=2}^N \frac{||{\bf x}_{1,j}(t) - {\bf x}_{1,1}(t)||}{N-1} dt,$$
$$E_{intra_2} = \lim\limits_{T\to \infty} \frac{1}{T} \int_{t_{trans}}^{t_{trans}+T} \sum\limits_{j=2}^N \frac{||{\bf x}_{2,j}(t) - {\bf x}_{2,1}(t)||}{N-1} dt,$$

and
\begin{equation}{\label{eq.9}}
E_{inter} = \lim_{T \to \infty} \frac{1}{T} \int_{t_{trans}}^{t_{trans}+T} \sum\limits_{j=1}^N \frac{||{\bf x}_{2,j}(t) - {\bf x}_{1,j}(t)||}{N} dt,
\end{equation}
where $\|\cdot\|$ signifies the Euclidean norm, $t_{trans}$ indicates the transient time of the simulation, and $T$ is a adequately large positive number. We consider the threshold of synchronization errors to be $10^{-5}$ to confirm the emergence of both synchronization. Using Runge--Kutta--Fehlberg algorithm the multiplex hypergraph (\ref{eq.4}) is solved numerically  with integration step $dt=0.01$. All the results are obtained by taking average over 10 network realizations.

In Figure \ref{fig3}(a), we have delineated the intralayer synchronization error by varying $\sigma$ for different values of $\lambda$, both for pairwise multiplex network and multiplex hypergraph. In both cases, a smooth change from the asynchronous intralayer state with $E_{intra}>0$ to an intralayer synchronous state where $E_{intra}=0$ is observed. For the pairwise multiplex network ( solid diamonds ), the threshold value of intralayer coupling is almost the same for different values of $\lambda$. For this case, the critical coupling is $\sigma \approx 0.092$. On the other hand, for multiplex hypergraph ( void circles ), the transition point from incoherent to coherent state is achieved at a comparably lower value of intralayer coupling strength, $\sigma \approx 0.079$. Also, in this case, the critical values are almost equal for various values of interlayer coupling strength.

\par Therefore, from the figure, it is conspicuous that consideration of higher-order structure in multiplex network enhances intralayer synchrony compared to corresponding pairwise multiplex network.      

\par Similarly, we acquire the interlayer synchronization error $E_{inter}$ for several values of intralayer coupling strength by varying the interlayer coupling parameter, depicted in Figure \ref{fig3}(b). For pairwise multiplex network ( solid diamonds ), as the intralayer coupling strength increases, the critical interlayer coupling for the transition from incoherent interlayer state to coherent state decreases up to the maximum intralayer coupling value $\sigma=0.05$. A similar transition occurs for the multiplex hypergraph too ( void circles ). So, in this case, unlike intralayer synchronization, no such interpretation can be established about the advancement of synchronization between counterpart nodes in the multiplex hypergraph realm as a function of interlayer coupling strength $\lambda$.       

\begin{figure*}[ht]
\centerline{\includegraphics[scale=0.4]{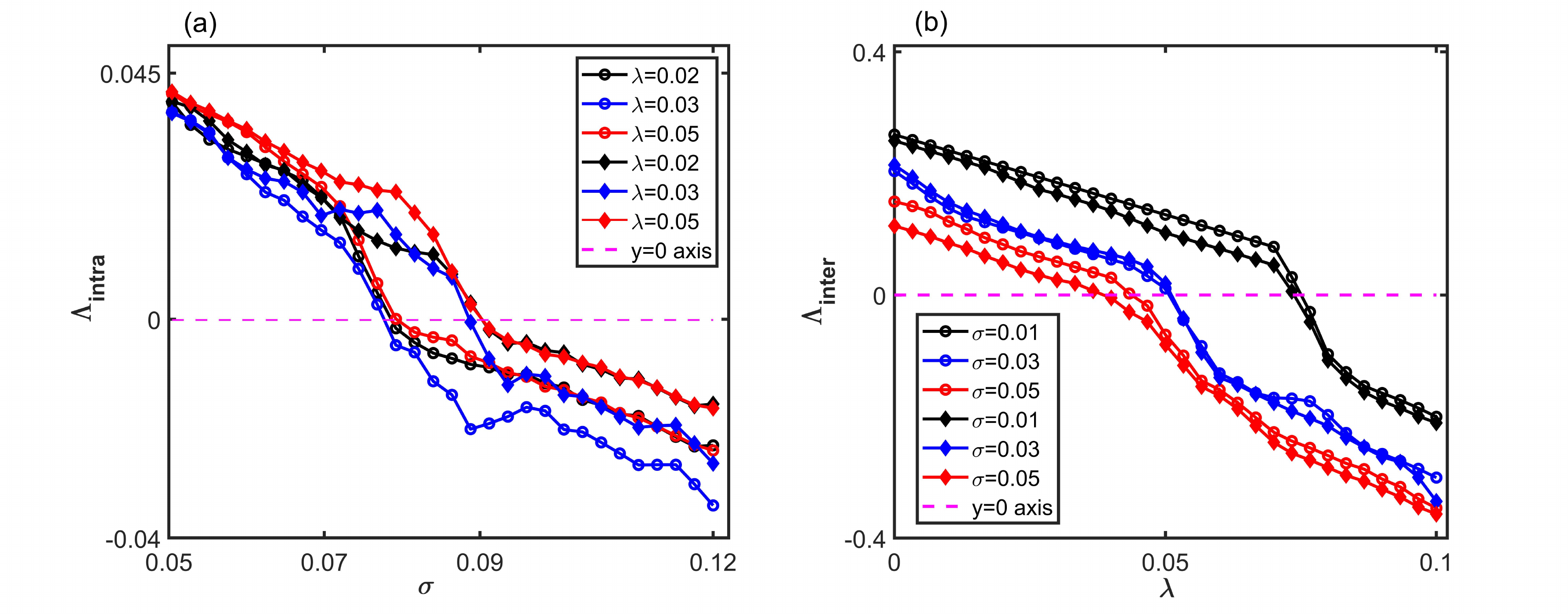}}
	\caption{ {\bf Variation of Maximum Lyapunov exponents.} (a) $\varLambda_{intra}$, maximum transverse Lyapunov exponent to intralayer synchronization manifold, as a function of $\sigma$ for various values of $\lambda$ same as Figure \ref{fig3} (a),  (b) $\varLambda_{inter}$, maximum transverse Lyapunov exponent to interlayer synchronization manifold, as a function of $\lambda$ for different values of $\sigma$ as in Figure \ref{fig3} (b). Solid diamonds are for the pairwise multiplex network, while void circles correspond to the multiplex hypergraph. The horizontal dashed magenta line corresponds to the zero level.} 
	\label{fig4}
\end{figure*}

\subsection{Linear stability analysis}

Now, we derive the necessary conditions for  stability of the two synchronization states in the multiplex hypergraph \eqref{eq.4} analytically using MSF approach. For intralayer synchronization, the oscillators of each layers are in synchrony, irrespective of synchronization between the replica nodes. Existence and invariance of intralayer synchronization solutions ${\bf{x}}_{1,S}={\bf{x}}_{1,i}$ and ${\bf{x}}_{2,S}={\bf{x}}_{2,i}$, for $i=1, 2, \cdots, N$ are guaranteed by the diffusive nature of the coupling function. In order to analyze the stability of intralayer synchronization solution, we consider small perturbations around the synchronization states, i.e., $\delta {\bf{x}}_{1,i}= {\bf{x}}_{1,i}-{\bf{x}}_{1,S}$ and $\delta {\bf{x}}_{2,i}= {\bf{x}}_{2,i}-{\bf{x}}_{2,S}$, and carry out linear stability analysis of Eqs. \eqref{eq.5} and \eqref{eq.6} . Then the linearized equation becomes, 
    
\begin{equation}\label{eq.10}
\begin{array}{l}
\delta\dot{\mathbf{x}}_{1,i}=JF(\mathbf{x}_{1,S})\delta\mathbf{x}_{1,i}-\sigma\sum\limits_{j=1}^{N}{{L}_{ij}^H}JG(\mathbf{x}_{1,S})\delta\mathbf{x}_{1,j}  \\~~~~~~ +\lambda  [JH(\mathbf{x}_{2,S})\delta{\mathbf x}_{2,i} - JH(\mathbf{x}_{1,S})\delta{\mathbf x}_{1,i})], \\\\

\delta\dot{\mathbf{x}}_{2,i}=JF(\mathbf{x}_{2,S})\delta\mathbf{x}_{2,i}-\sigma\sum\limits_{j=1}^{N}{{L}_{ij}^H}JG(\mathbf{x}_{2,S})\delta\mathbf{x}_{2,j}  \\~~~~~~ +\lambda  [JH(\mathbf{x}_{1,S})\delta{\mathbf x}_{1,i} - JH(\mathbf{x}_{2,S})\delta{\mathbf x}_{2,i}].
\end{array}
\end{equation}
Here $J$ is the Jacobian operator and the synchronized solutions $({\bf{x}}_{1,S}, {\bf{x}}_{2,S})$ satisfies,

\begin{equation}{\label{eq.11}}
\begin{array}{lcl}
\dot{\bf x}_{1,S} = F({\bf x}_{1,S}) + \lambda [H({\bf x}_{2,S}) - H({\bf x}_{1,S})], \\\\
\dot{\bf x}_{2,S} = F({\bf x}_{2,S}) + \lambda [H({\bf x}_{1,S}) - H({\bf x}_{2,S})].
\end{array}
\end{equation}

Now we rewrite the \eqref{eq.10} in block matrix form introducing variational vector 
$\delta {\bf x}_1= [ \delta {\bf x}_{1,1}^{tr}, \delta {\bf x}_{1,2}^{tr}, \cdots, \delta {\bf x}_{1,N}^{tr}]^{tr}$ and $\delta {\bf x}_2= [ \delta {\bf x}_{2,1}^{tr}, \delta {\bf x}_{2,2}^{tr}, \cdots, \delta {\bf x}_{2,N}^{tr}]^{tr}$ as, 

\begin{equation}{\label{eq.12}}
\begin{array}{l}
\delta\dot{\mathbf{x}}_{1}=[I_{N} \otimes JF(\mathbf{x}_{1,S})-\sigma \mathscr{L}^{H}\otimes JG]\delta{\mathbf{x}}_{1} \\ ~~~~~
+\lambda [I_{N}\otimes JH(\mathbf{x}_{2,S})\delta\mathbf{x}_{2}-I_{N}\otimes JH(\mathbf{x}_{1,S})\delta\mathbf{x}_{1}], \\\\

\delta\dot{\mathbf{x}}_{2}=[I_{N} \otimes JF(\mathbf{x}_{2,S})-\sigma \mathscr{L}^{H}\otimes JG]\delta{\mathbf{x}}_{2} \\ ~~~~~
+\lambda [I_{N}\otimes JH(\mathbf{x}_{1,S})\delta\mathbf{x}_{1}-I_{N}\otimes JH(\mathbf{x}_{2,S})\delta\mathbf{x}_{2}].

\end{array}
\end{equation}

Since, $L^{H}$ is real symmetric zero row-sum matrix,\cite{random_pre} it is diagonalizable and all its eigenvalues $\gamma_{i}$ $(i=1, 2, \cdots, N)$ are non-zero real numbers with smallest eigenvalue zero. The corresponding set of eigenvectors forms an orthogonal basis of $\mathbb{R}$. The variational equation \eqref{eq.12} contains all the parallel and transverse components to the synchronization manifold. To decouple the transverse modes from parallel one, we project the stack variables $\delta {\bf x}_{l}$ $(l=1,2)$ onto the basis of eigenvector $V=[v_1, v_2, \cdots, v_N]$ corresponding to the Laplacian $L^{H}$ by defining new variables, ${\mathbf{\eta}}^{(l)}=(V \otimes I_{d})^{-1}\delta {\bf x}_{l}$, where $v_1=\frac{1}{\sqrt{N}}(1, 1, \cdots, 1)^{tr}$ is the eigenvector corresponding to $\gamma_{1}=0$. Then, the dynamics of  variational equation in terms of the new variable becomes,

\begin{equation}{\label{eq.13}}
\begin{array}{l}
\dot{\mathbf{\eta}}^{(1)}=[I_{N} \otimes JF(\mathbf{x}_{1,S})-\sigma {\Gamma}\otimes JG]{\mathbf{\eta}}^{(1)} \\ ~~~~~
+\lambda [I_{N}\otimes JH(\mathbf{x}_{2,S}){\mathbf{\eta}}^{(2)}-I_{N}\otimes JH(\mathbf{x}_{1,S}){\mathbf{\eta}}^{(1)}], \\\\

\dot{\mathbf{\eta}}^{(2)}=[I_{N} \otimes JF(\mathbf{x}_{1,S})-\sigma {\Gamma}\otimes JG]{\mathbf{\eta}}^{(2)} \\ ~~~~~
+\lambda [I_{N}\otimes JH(\mathbf{x}_{2,S}){\mathbf{\eta}}^{(1)}-I_{N}\otimes JH(\mathbf{x}_{1,S}){\mathbf{\eta}}^{(2)}],

\end{array}
\end{equation}
where
\begin{equation*}
V^{-1}{L}^{H}V=diag\{0=\gamma_{1}<=\gamma_{2}<=\cdots<=\gamma_{N}\}=\Gamma. 
\end{equation*} 

We can now decouple the dynamics of the linearized system into $N$ numbers of $2d$ dimensional equations given by,
\begin{equation}{\label{eq.14}}
\begin{array}{l}
\dot{\mathbf{\eta}}^{(1)}_{i}=JF(\mathbf{x}_{1,S})\mathbf{\eta}^{(1)}_{i}-\sigma \gamma_{i}JG{\mathbf{\eta}}^{(1)}_{i} \\ ~~~~~
+\lambda [JH(\mathbf{x}_{2,S}){\mathbf{\eta}}^{(2)}_{i}- JH(\mathbf{x}_{1,S}){\mathbf{\eta}}^{(1)}_{i}], \\\\

\dot{\mathbf{\eta}}^{(2)}_{i}=JF(\mathbf{x}_{2,S})\mathbf{\eta}^{(2)}_{i}-\sigma \gamma_{i}JG{\mathbf{\eta}}^{(2)}_{i} \\ ~~~~~
+\lambda [JH(\mathbf{x}_{1,S}){\mathbf{\eta}}^{(1)}_{i}- JH(\mathbf{x}_{2,S}){\mathbf{\eta}}^{(2)}_{i}]. 

\end{array}
\end{equation}

The dynamics of $\eta^{(l)}_{1}$ represents the motion along intralayer synchronized manifold and that for the other states $\eta^{(l)}_{i}$, $i=2, \cdots, N$ accounts for different modes transverse to the synchronization manifold. The problem of stability is then brought down to solve the nonlinear equation \eqref{eq.11} along with the linear equation \eqref{eq.14} for the calculation of maximum transverse Lyapunov exponents. We consider $\varLambda_{max}^1$ and $\varLambda_{max}^2$ are the maximum transverse Lyapunov exponents associated with the two layers. The necessary condition for the synchronization solution requires $\varLambda_{intra} = \max\{\varLambda_{max}^1,\varLambda_{max}^2\}$ to be negative, .i.e., all the transverse modes to be die out. Given the node dynamics and coupling functions, $\varLambda_{intra}$ is mainly function of the two coupling parameters and the eigenvalues of Laplacian matrix, $L^H$, i.e., $\varLambda_{intra}= \varLambda_{intra} (\sigma, \lambda, \Gamma)$. To validate the analytical condition, we portray $\varLambda_{intra}$ by varying $\sigma$ in Figure \ref{fig4}(a) for the identical values of $\lambda$ described in Figure \ref{fig3}(a). For both the multiplex hypergraph and pairwise multiplex network, the sign of maximum Lyapunov exponents $\varLambda_{intra}$ switches from positive to negative precisely at the identical values of $\sigma$ where $E_{intra}$ achieves the threshold corresponding to intralayer synchronization. Thus our analytical Master Stability function approach nicely upholds the numerical results and supports the enhancement in the multiplex hypergraph realm.

\par In addition, the stability of interlayer synchronization state $\mathbf{x}_1=\mathbf{x}_2=\mathbf{x}$ is also investigated, whose existence and invariance is guaranteed as oscillators between the layers are diffusively coupled. To do this, we consider small perturbations $\delta{\mathbf{x}}_{i}= \mathbf{x}_{2,i}-\mathbf{x}_{1,i}$ around the synchronization solution, and execute linear stability analysis. We get the $N\times d$ linearized equations satisfying,

\begin{equation}\label{eq.15}
\begin{array}{l}
    \delta\dot{\mathbf{x}}_{i}=JF(\mathbf{x}_{i})\delta\mathbf{x}_{i}-\sigma\sum\limits_{j=1}^{N}{L}^{H}JG(\mathbf{x}_{i})\delta\mathbf{x}_{j}-2 \lambda JH(\mathbf{x}_{i}) \delta\mathbf{x}_{i}, 
\end{array}
\end{equation}
where $\mathbf{x}$ is the state variable corresponding to interlayer synchronization solution satisfying,

\begin{equation}\label{eq.16}
\begin{array}{l}
 \dot{\mathbf{x}}_{i}=F(\mathbf{x}_{i})-\sigma \sum \limits_{j=1}^{N} {L}^{H}_{ij} G \mathbf{x}_{j}.
\end{array}
\end{equation}

The linearized equation \eqref{eq.15} is solved together with $N\times d$ nonlinear equation \eqref{eq.16} to calculate all transverse Lyapunov exponents associated with the interlayer synchronization manifold. For the coherent interlayer  state to be stable, the necessary condition requires maximum transverse Lyapunov exponent $\varLambda_{inter}$ to be negative as we vary the coupling strengths, i.e., when all the transverse modes to the synchronization manifold become extinct. Figure \ref{fig4}(b) displays the variation of $\varLambda_{inter}$ with respect to interlayer coupling $\lambda$ for different values of $\sigma$ as in Figure \ref{fig3}(b). The $\varLambda_{inter}$ curves cross the zero axis and become negative at the exact values of $\lambda$ stated before in the previous section. Nevertheless, these curves pass over zero for both multiplex hypergraph and pairwise multiplex network almost at the same values of $\sigma$, which perfectly endorse our numerical results presented in Figure \ref{fig3}(b). 

\subsection{Spectral Interpretation} Other than linear stability analysis, the spectral analysis of the Laplacian matrices is interpreted for an acute understanding of the advancement of synchronization in the multiplex hypergraph regime. For a network with an unbounded region of synchronization,\cite{boccaletti2006complex}  i.e., with type-II MSF, synchronizability depends on the spectral gap $\gamma_{2}$ (smallest non-zero eigenvalue of the corresponding Laplacian matrix).

\begin{figure}[ht]
	\centerline{\includegraphics[scale=0.08]{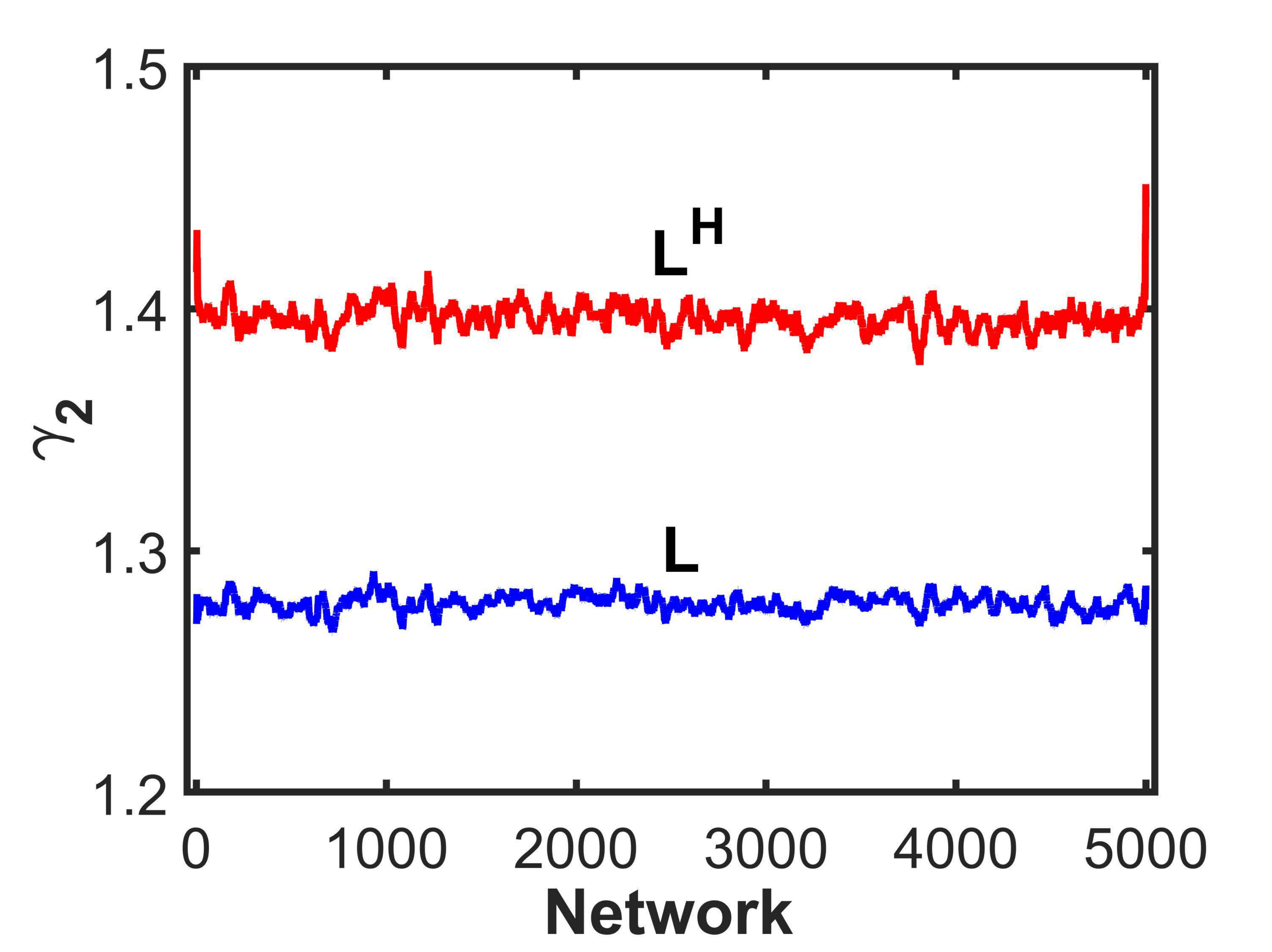}}
		\caption{{\bf Spectral gap of 5000 sampled scale-free networks with $N=500$ nodes.} Blue and Red markers represent the spectral gap $(\gamma_{2})$ corresponding to the Laplacian of pairwise network and hypergraph, respectively. } 
	\label{fig5}
\end{figure}

A larger value of $\gamma_{2}$ indicates a better synchronizability, that is to say, a smaller coupling strength is needed to achieve synchronization. Here we generate 5000 sample networks with $N=500$ nodes in each layer, constructed using the mechanism illustrated before. For each of these networks, we have estimated the synchronizability measure $\gamma_{2}$ based on both hypergraph and pairwise network Laplacians $L$ and $L^{H}$, respectively. Figure \ref{fig5} shows no cases where the spectral gap of hypergraph Laplacian is smaller than that of pairwise network Laplacian.  It affirms our analytical prediction that connectivity through higher-order interactions enriches the synchronization with respect to the pairwise connection.  

\begin{figure*}[ht]
	\centerline{\includegraphics[scale=0.6]{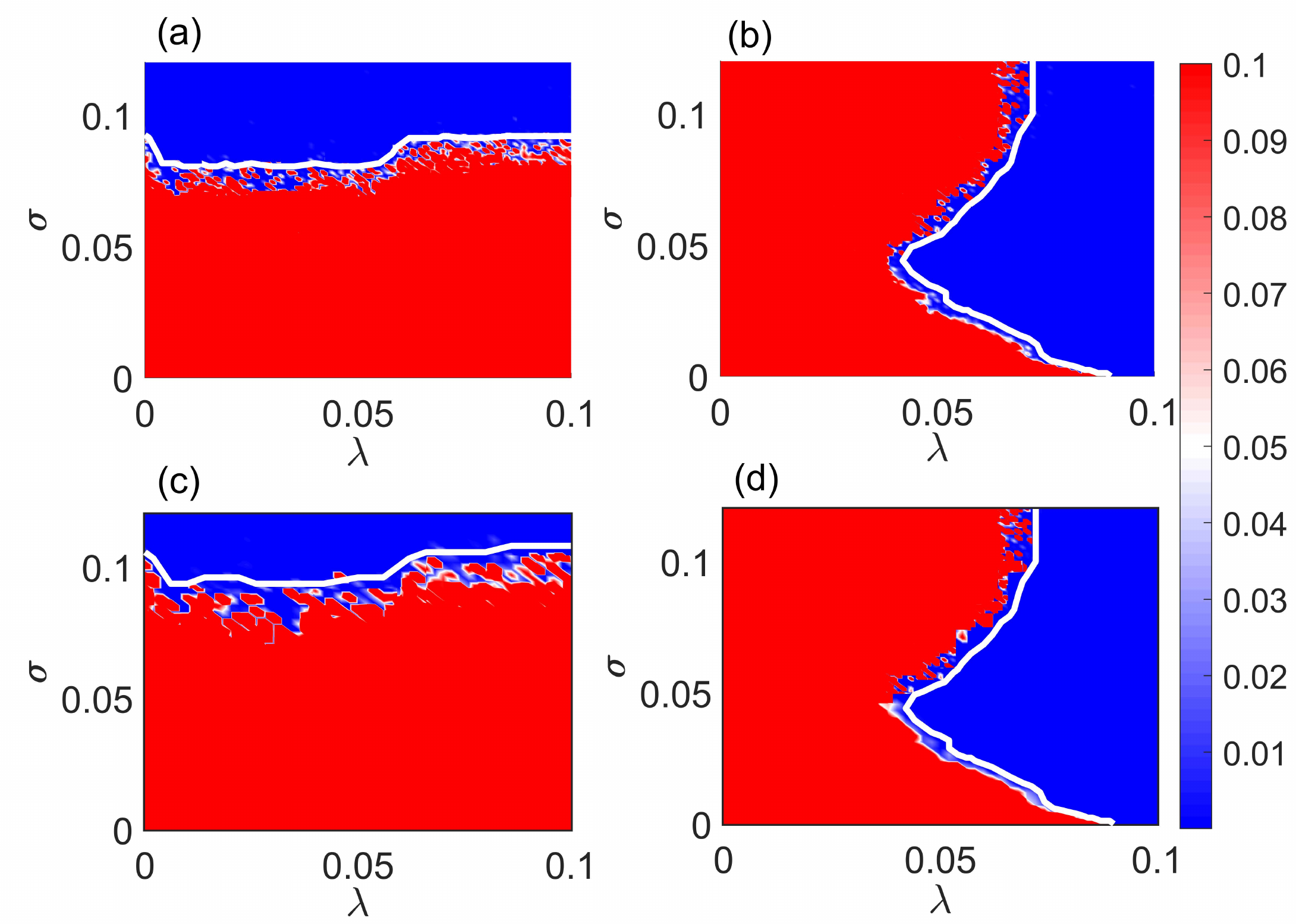}}
		\caption{{\bf Contour plots of synchronization errors for multiplex hypergraph and pairwise multiplex network.} (a) and (c) represents $E_{intra}$ in $(\lambda, \sigma)$ parameter space for multiplex hypergraph and pairwise multiplex network, respectively. Horizontal solid white curve corresponds to the theoretical bound for intralayer synchronization obtained from Eq. \eqref{eq.14}, and (b), (d) delineate $E_{inter}$ in $(\lambda, \sigma)$ plane for multiplex hypergraph and pairwise multiplex network, respectively . The vertical solid white curve corresponds to the analytical prediction for interlayer synchronization threshold obtained from Eq. \eqref{eq.15}. In each panel red region indicates the region of desynchronization, and the blue region corresponds to a stable synchronization state.} 
	\label{fig6}
\end{figure*}

\par Besides, we inspect the intralayer and interlayer synchronization for multiplex hypergraph by varying both the coupling strength in $(\lambda, \sigma)$ parameter plane and compared the result with the pairwise multiplex network. Figures \ref{fig6}(a) and \ref{fig6}(b) display the complete scenario of the intralayer and interlayer synchronizations, respectively for multiplex hypergraph. The coherence (deep blue) and incoherent (deep red) region in both subfigures is separated by solid white lines, which are theoretical conjectures of synchronization thresholds acquired from equations \eqref{eq.14} and \eqref{eq.15}. It is perceptible from Figure \ref{fig6}(a) that for varying intralayer coupling strength, intralayer synchronization is only slightly affected by the appearance of interlayer coupling $\lambda$. The threshold for interlayer synchrony (Figure \ref{fig6}(b)) decreases as the value of $\sigma$ increases up to $\sigma \approx 0.052$. Beyond that, as the value of $\sigma$ increases, the interlayer synchronization threshold moves toward a higher value of $\lambda$ until the intralayer synchronization is achieved at $\sigma \approx 0.0912$. Further increment in the value of $\sigma$ does not affect the threshold of interlayer synchronization. Hence, Figure \ref{fig6}(b) indicates that the interlayer synchrony is always discerned for large values of $\lambda$ irrespective of intralayer synchronization. However, for the pairwise multiplex network, the contour plots of intralayer and interlayer synchronization is depicted in Figures \ref{fig6}(c) and \ref{fig6}(d). In pairwise multiplex network too, both intralayer and interlayer synchronization shows almost similar behavior with respect to varying coupling strengths. But for the pairwise multiplex network scenario, the threshold value for intralayer synchronization is higher compared to the multiplex hypergraph case. Thus rigorously plotting intralayer and interlayer synchronization regions for varying coupling strengths, we can reassert our result that intralayer synchronization is enhanced by introducing higher-order structure in the multiplex network.  At the same time, no such interpretation can be drawn for the interlayer synchronization phenomena.

\subsection{Robustness of Interlayer synchronization}

\begin{figure}[ht]
\centerline{\includegraphics[width=8.5cm, height=6cm]{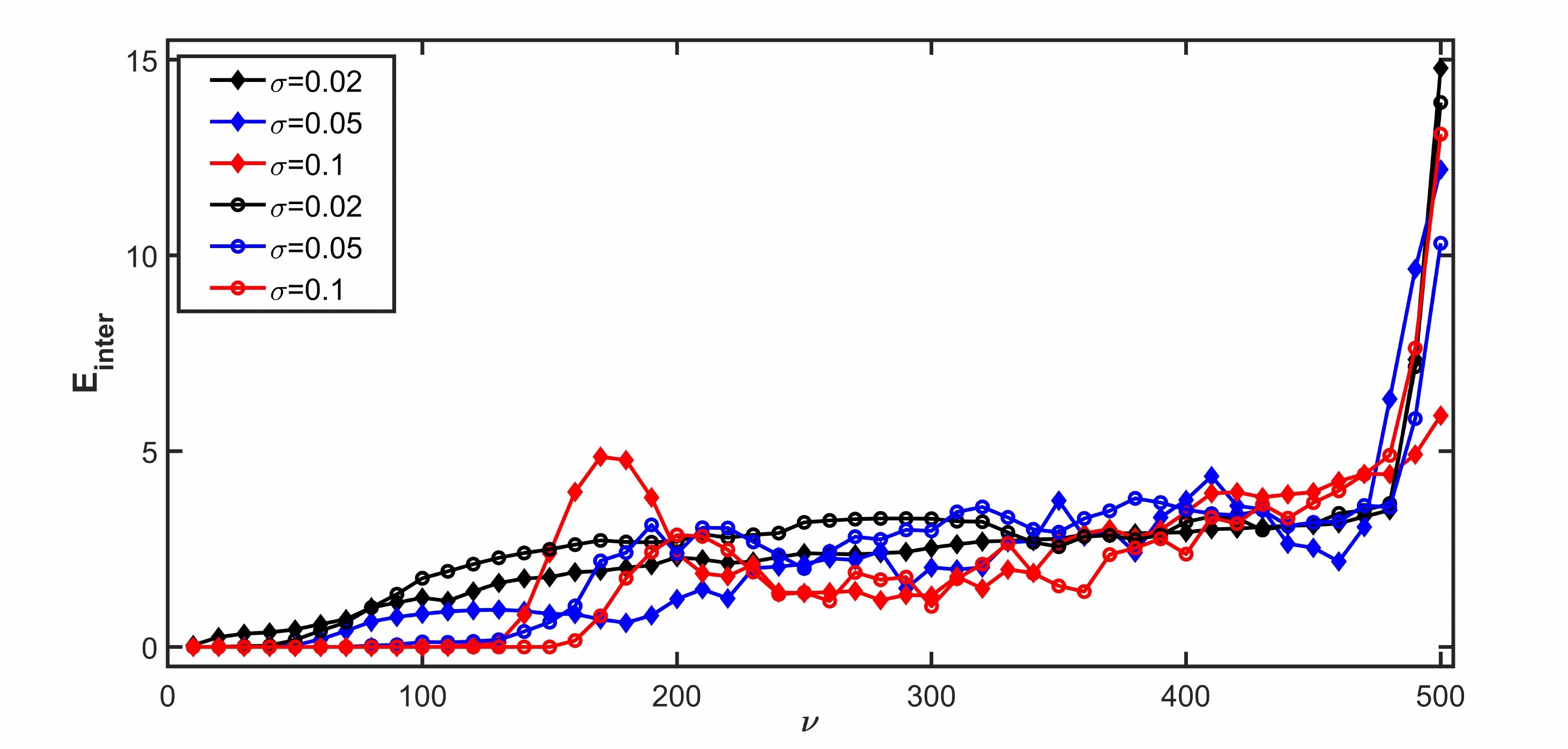}}
		\caption{ {\bf Persistence of interlayer synchronization.} $E_{iner}$ as a function of demultiplexed nodes $(\nu)$ for a fixed value of $\lambda=0.1$ and increasing values of $\sigma$ (see in the legend). Solid diamonds represent the pairwise multiplex network and void circles correspond to the multiplex hypergraph.} 
	\label{fig7}
\end{figure}

Lastly, we explore the endurance of interlayer synchronization against gradual random deportation of links connecting the replica nodes of the two layers, called demultiplexing of the multiplex network. We sequentially remove links connecting replica nodes of the multiplex network until both the layers become isolated. Starting from a stage of complete interlayer synchrony for a fixed interlayer coupling $\lambda=0.1$, we evaluate the synchronization error $E_{inter}$ as a function of the number of demultiplexed nodes $\nu$ for increasing values of intralayer coupling strength $\sigma$. Figure \ref{fig7} displays the results for multiplexes with hypergraph (void circles) and pairwise network (solid diamonds) structure. For $\sigma=0.02$ with the pairwise multiplex network configuration, synchronization error $E_{inter}$ becomes non-zero as soon as the first pair of replica nodes is disconnected.

 On contrary for the multiplex hypergraph, interlayer synchronization persists up to the removal of $\nu=40$ nodes. As the intralayer coupling strength $\sigma$ increases to 0.05, the interlayer synchrony perseveres up to the removal of 60 interlayer links for the pairwise multiplex network. In the case of multiplex hypergraph structure, synchronization is more robust as earlier, $E_{inter}$ remains zero till extraction of 110 links between replica nodes. Further increment of $\sigma$ shows a similar transition scenario. Thus, we can conclude that the interlayer synchrony is more robust when we consider the higher-order structures in the multiplex framework. However, interlayer synchronization becomes more persistent with increasing intralayer coupling strength for both cases.

\section{Conclusion}
 Summing up, we have here explicitly scrutinized the two fundamental synchronization scenarios, i.e., intralayer and interlayer synchronization in a multiplex network with higher-order structure, schematized by hyperedges. The intralayer connections are constructed from a scale-free network by converting the maximal cliques of sizes $p$ to hyperedges of size $p$ while the interlayer connections are pairwise, representing only the binary interaction between replica nodes. Using linear stability analysis, the necessary circumstances for the stability of the two synchronization states are derived analytically. The study suggests that the intralayer synchronization enhances when we consider beyond pairwise interactions in multiplex structure compared to the multiplex networks with pairwise interaction. In contrast, the threshold for interlayer synchrony is almost the same for both two incidents. Further, we also addressed the tenacity of the interlayer synchronization against successive removal of connections between counterpart nodes. Notably, the multiplex network with higher-order structure, i.e., multiplex hypergraph shows more robust interlayer synchrony than the multiplex network with only pairwise interactions. So introducing only higher-order interactions through hyperedges instead of pairwise connections can enhance the intralayer synchrony and allow interlayer synchrony to be more persistent  for scale-free network topology. We believe that the similar results may be verified by considering other types of networks (e.g., random network, small-world networks etc.).  We expect that our study will facilitate several new perceptions of collective behaviors in the multilayer realm.

\section*{DATA AVAILABILITY}
The data that support the findings of this study are available within the article.

\end{document}